# Investigating Energy-Dependent Anisotropy in Cosmic Rays with IceTop Surface Array

**The IceCube Collaboration**
(a complete list of authors can be found at the end of the proceedings)

*E-mail:* rabbasi@luc.edu, paolo.desiati@icecube.wisc.edu, juancarlos@icecube.wisc.edu, mcnally_ft@mercer.edu

This study presents preliminary results from the analysis of cosmic-ray anisotropy using air showers detected by the IceTop surface array between 2011 and 2022. With improved statistical precision and updated Monte Carlo simulation events compared to previous IceTop reports, we investigate anisotropy patterns across four energy ranges spanning from 300 TeV to 6.9 PeV. This work extends the measurement of cosmic-ray anisotropy in the southern hemisphere to higher energies than previously achieved with IceTop. Our results provide a foundation for exploring potential connections between the observed anisotropy, the energy spectrum, and the mass composition of the cosmic-ray flux.

**Corresponding authors:**
R. Abbasi[1*], P. Desiati[2], Juan Carlos Díaz Vélez[2], F. McNally[3]
[1] *Loyola University Chicago, Chicago, USA*
[2] *Dept. of Physics and Wisconsin IceCube Particle Astrophysics Center, University of Wisconsin—Madison, Madison, WI 53706, USA*
[3] *Mercer University, Georgia, USA*

[*] *Presenter*

## 1. Introduction

In the TeV to PeV energy range, the distribution of cosmic-ray arrival directions exhibits measurable deviations from isotropy on large angular scales. These anisotropies, typically at the level of $10^{-3}$ or less, are thought to result from the interplay between cosmic-ray source distributions, diffusive propagation through the turbulent Galactic magnetic field, and possible local magnetic structures.

Despite the high degree of isotropy expected from diffusive transport over kiloparsec scales, persistent large-scale patterns—such as dipole and quadrupole components—have been detected by multiple ground-based observatories, including the Tibet AS$\gamma$ [1], Super-Kamiokande [2], Milagro [3], EAS-TOP [4], KASCADE-Grande [5], MINOS [6], and HAWC [7] experiments in the Northern Hemisphere and IceCube [8–12], including its surface air-shower array IceTop [13], in the Southern Hemisphere. These features provide critical constraints on models of galactic cosmic-ray transport and the structure of the local interstellar magnetic field. Studying both the spatial and energy-dependent characteristics of these anisotropies is key to understanding the transition between local and global contributions to the observed cosmic-ray flux, as energy-dependent patterns can reflect changes in propagation regimes, source distributions, and the influence of magnetic field structures at different scales.

Located at the geographic South Pole, the IceCube Observatory is currently the only experiment capable of reporting on the cosmic-ray anisotropy in the Southern Hemisphere in the TeV–PeV energy region. The IceTop surface array, situated at an altitude of 2835 m on the Antarctic ice sheet directly above IceCube's in-ice component, is a dedicated cosmic-ray air shower array optimized for PeV-scale observations. Unlike the in-ice detector, it detects both the hadronic and electromagnetic components of extensive air showers at ground level.

Earlier studies of large-scale anisotropy using IceTop were based on data from partial detector configurations [13]. In this work, we present updated results using data collected by the complete IceTop array between May 2011 and May 2022. This extended dataset significantly reduces statistical uncertainties. In addition, it allows for more robust estimates of systematic uncertainties. The use of data from a stable, fully operational detector combined with improved analysis techniques and continuous, year-round coverage—makes it possible to probe anisotropy in the PeV energy range with minimal systematic uncertainties.

## 2. Energy Binning and Event Selection Criteria

The IceTop array completed its installation of 81 stations in 2011. These stations are distributed over an area of 1 km$^2$ in a hexagonal grid, with a spacing of approximately 125 m between neighboring stations. Each IceTop station consists of two tanks, and each tank houses two Digital Optical Modules (DOMs) with different photomultiplier tube (PMT) gains. For an event to be registered, at least six DOMs—corresponding to three stations—must trigger within a 5 $\mu$s time window. A detailed description of the IceTop detector can be found in [14]. IceTop detects air showers generated by cosmic-rays particles with energies exceeding 300 TeV [15].

Studies of the arrival direction distribution of cosmic rays in the TeV to PeV energy range have revealed a variation with energy in both the amplitude and phase of the large-scale anisotropy [2, 3, 5, 13, 16–24]. To investigate this energy dependence, we split the dataset into four energy bands, using the number of triggered IceTop stations as a proxy for energy.

A critical consideration when using the number of IceTop stations as an energy proxy is the snow accumulation at the South Pole. This accumulation, driven by wind, affects the surface detector's performance. As a result, the relationship between the number of triggered stations and the cosmic-ray energy changes over time. For example, a five-station triggered event in 2011 corresponds to a lower energy than a five-station triggered event in 2021.

To ensure consistent median energies across energy bands over time, we adjust the station-count cuts by approximately one station every two years. For instance, an energy band defined by triggering 10–17 stations in 2011 would correspond to 9–16 stations in 2013, and so on, up to 2021. This strategy maintains stable median energies for each band (referred to as "Tiers") over the full dataset. The stability of these energy bands was validated by confirming that the S125 parameter distributions remained consistent across years within each Tier. S125 is a measure of the shower size at 125 m from the shower axis and serves as an energy proxy. These adjustments and validations were developed using a 10% burn sample of the data to avoid biasing the final results; this subset allows for method development while preserving the integrity of the full dataset for the final analysis.

To maintain uniform median energy levels and ensure consistent resolution across the anisotropy sky maps, we investigated the median energy as a function of reconstructed zenith angle for each energy band, considering both proton and iron cosmic-ray primaries. As shown in Fig. 1, the median energy increases and becomes less consistent across the sky at larger zenith angles. In addition, the accuracy of the angular reconstruction deteriorates with increasing zenith angle. To preserve both energy and angular resolution, we apply a quality cut that excludes events with reconstructed zenith angles greater than 55°.

The primary particles' energy distribution for each energy band is shown in Fig. 2. Table 1 presents the median energies and the 68% containment intervals for proton and iron primaries, as well as for the all-particle spectrum. The all-particle spectrum is constructed as a weighted sum of four primary components—proton, helium, oxygen, and iron—according to the H4a flux model [25]. All simulation datasets are based on the hadronic interaction model Sibyll 2.1 [26].

| | Tier 1 | | Tier 2 | | Tier 3 | | Tier 4 | |
|---|---|---|---|---|---|---|---|---|
| | Median | 68% Interval | Median | 68% Interval | Median | 68% Interval | Median | 68% Interval |
| Proton | 0.26 PeV | 0.14 – 0.53 PeV | 0.77 PeV | 0.37 – 1.55 PeV | 1.93 PeV | 1.10 – 3.70 PeV | 5.66 PeV | 3.41 – 12.06 PeV |
| Iron | 0.46 PeV | 0.23 – 0.98 PeV | 1.36 PeV | 0.70 – 2.98 PeV | 3.20 PeV | 1.78 – 7.19 PeV | 9.36 PeV | 4.89 – 23.90 PeV |
| All Particles | **0.30 PeV** | 0.15 – 0.63 PeV | **0.92 PeV** | 0.44 – 1.95 PeV | **2.36 PeV** | 1.29 – 4.75 PeV | **6.89 PeV** | 3.92 – 16.00 PeV |

**Table 1:** Median energy and 68% intervals for proton, iron, and all particles across four energy tiers.

## 3. Data Analysis and Observations

To study large-scale cosmic-ray anisotropy, we begin by pixelating the sky using the HEALPix library [27], which provides an equal-area binning of the celestial sphere, enabling consistent comparisons across the celestial map.

Given that the anisotropy under investigation is on the order of $10^{-3}$, and that detector simulations cannot accurately reproduce rate variations as a function of time and viewing angle, data-driven techniques are typically used to estimate the background. In this work, we rely on the likelihood-based reconstruction method described in [28] and applied in similar studies, including [22].

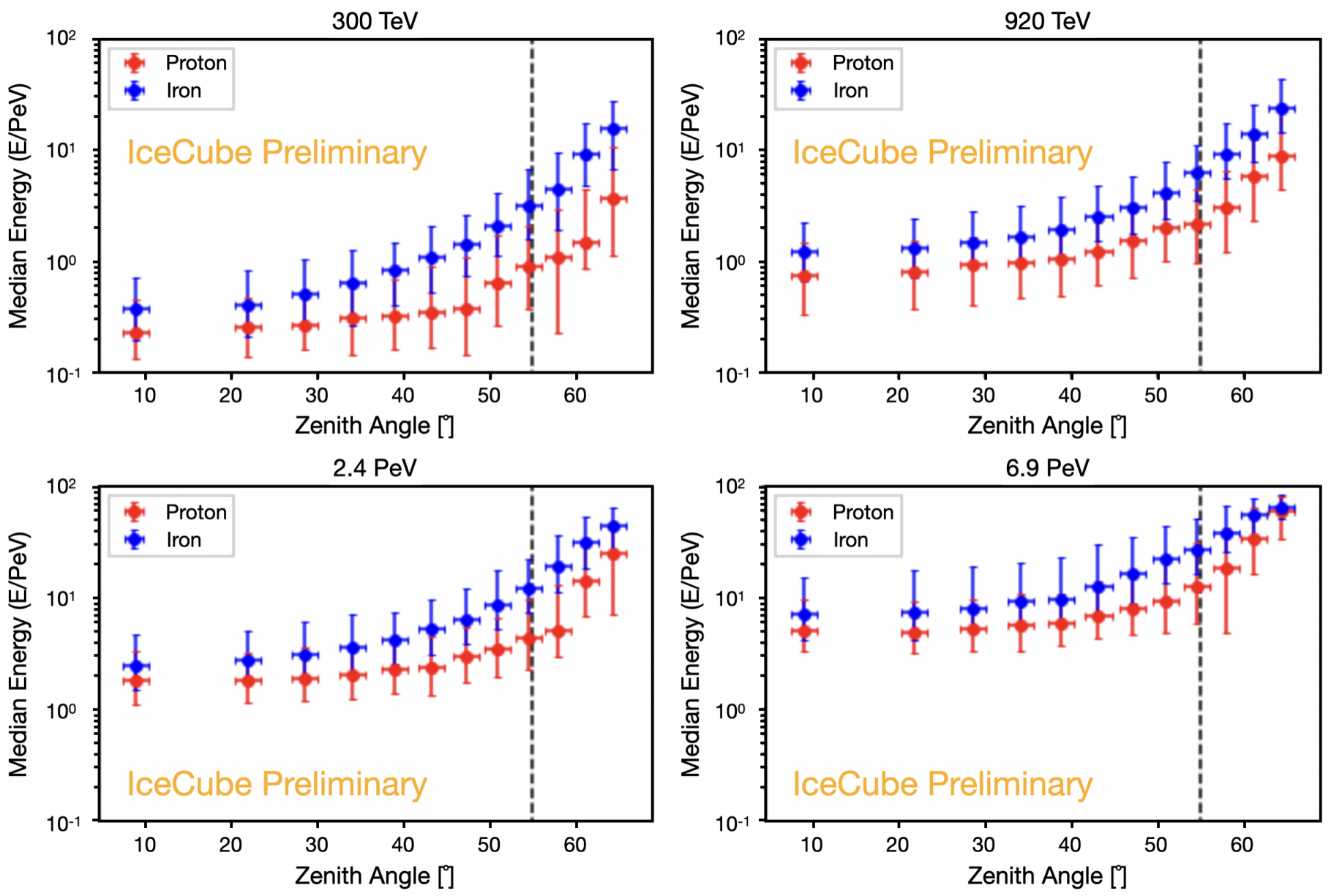

**Figure 1:** True primary median energy as a function of reconstructed zenith angle for the four selected energy bands for proton and iron cosmic-ray primaries. The error bars correspond to a 68% confidence interval. The dashed black line marks the zenith angle cut, restricting the analysis to events with reconstructed zenith angles less than 55 degrees.

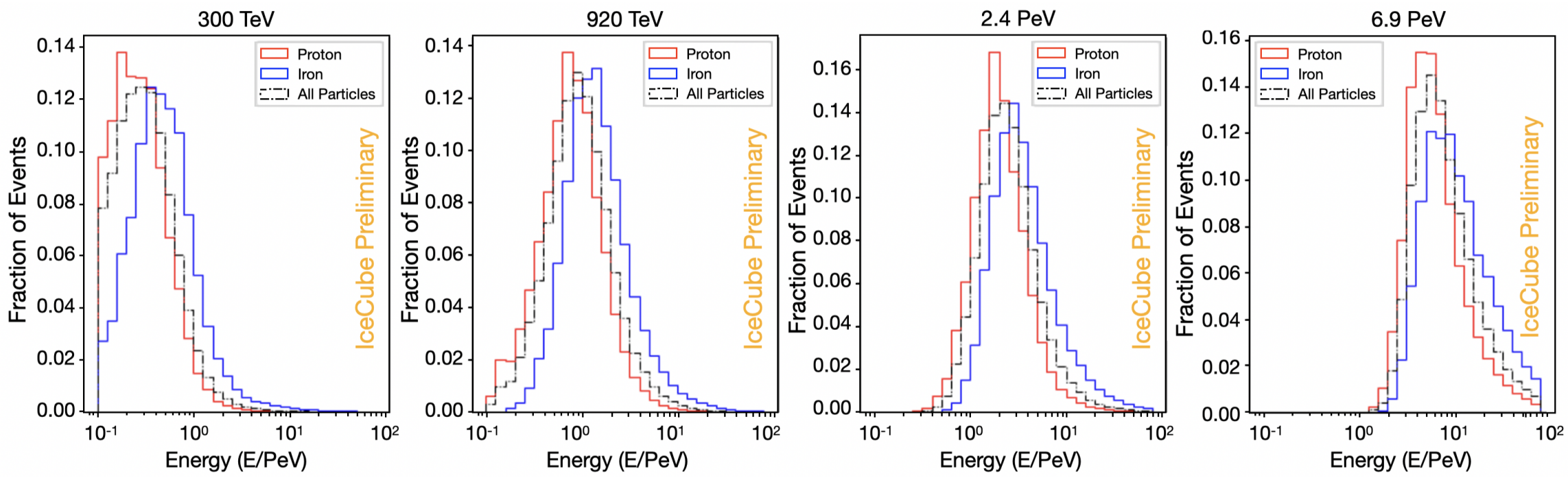

**Figure 2:** Simulated true energy distributions for events in the four selected energy bands, assuming iron (blue), proton (red), and all-component (dashed) compositions.

Once the background sky maps are generated (representing the response function to an isotropic cosmic-ray flux), they are compared with the data maps to produce relative intensity sky maps, which reveal deviations from isotropy. The relative intensity $\delta I$ in each pixel $i$ is calculated as:

$$\delta I_i = \frac{N_i - \langle N_i \rangle}{\langle N_i \rangle} \tag{1}$$

where $N_i$ is the number of data observed per pixel, and $\langle N_i \rangle$ is the expected number of background events. In addition, we provide statistical significance sky maps using the Li & Ma method [29].

A 20° radius top-hat smoothing method is applied, summing contributions from all surrounding pixels within that radius. This smoothing scale was chosen following the precedent set by previous large-scale anisotropy studies [12, 13], as it maximizes statistical significance while preserving sensitivity to broad structural features in the sky.

Figure 3 shows the relative intensity and statistical significance sky maps in J2000 equatorial coordinates for all four energy tiers. These maps are dominated by a broad deficit region between 60° and 120° in right ascension. To quantify the anisotropy, we project the two-dimensional sky maps onto one dimension by plotting the relative intensity as a function of right ascension (see Fig. 4).

Because the observed anisotropy manifests primarily as a deficit, a simple dipole model is inadequate. Instead, we fit the right ascension profiles using a third-order harmonic function:

$$\delta I(\alpha) = \sum_{n=1}^{3} A_n \cos[n(\alpha - \phi_n)] + B \tag{2}$$

where $A_n$ is the amplitude, $\alpha$ is the right ascension, $\phi_n$ is the phase of the $n^{\text{th}}$ harmonic component, and $B$ is a constant. The observed deficit is consistent with previous IceTop results in the ~300 TeV to ~1 PeV energy range [13]. Both the amplitude and width of the deficit increase across this range, while its location in right ascension remains relatively stable. At higher energies (2.4 and 6.9 PeV), the amplitude of the anisotropy remains comparable to that at 1 PeV, though the width of the deficit feature broadens further.

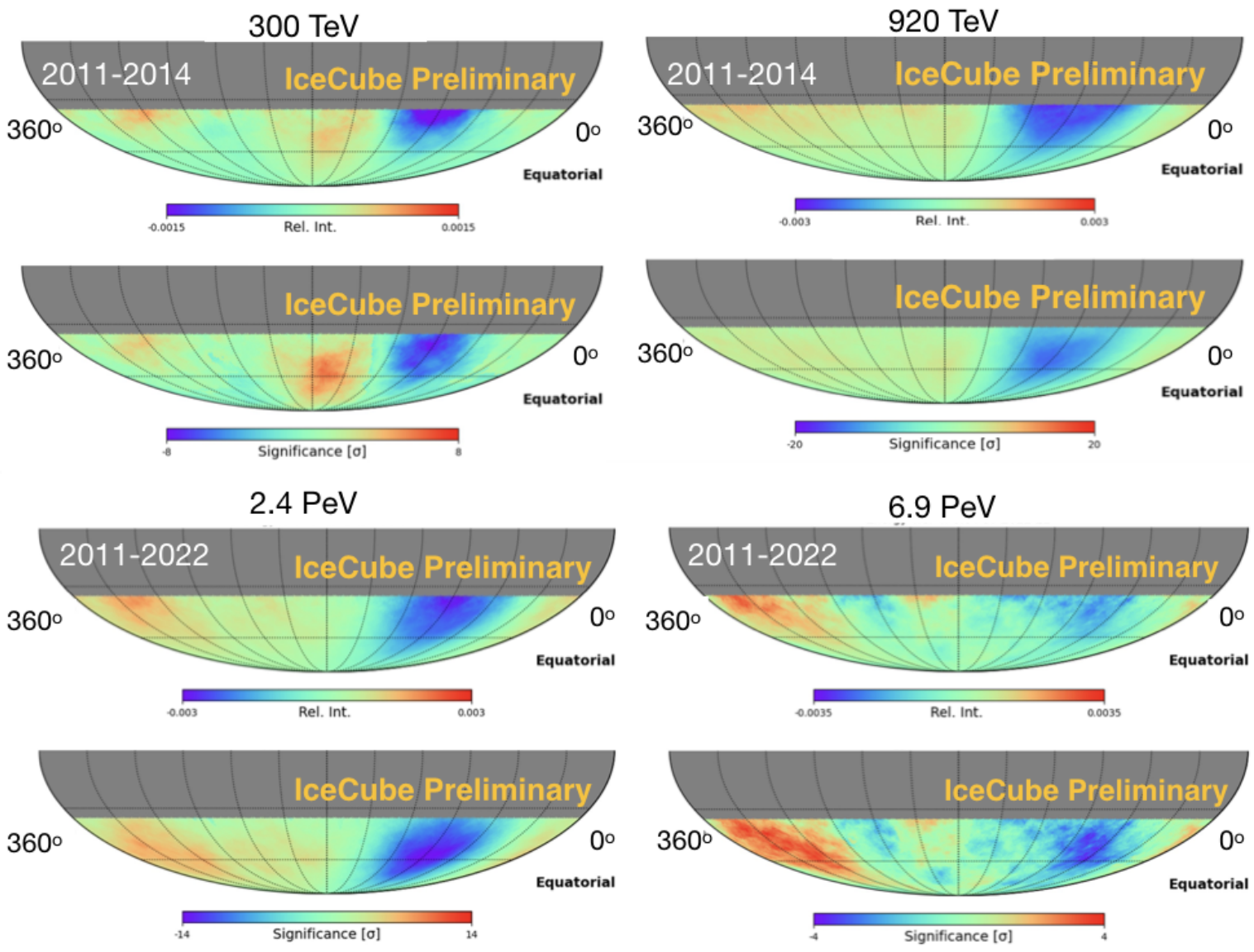

**Figure 3:** Relative intensity (*top*) and statistical significance (*bottom*) maps for four tiers with their corresponding median energies indicated.

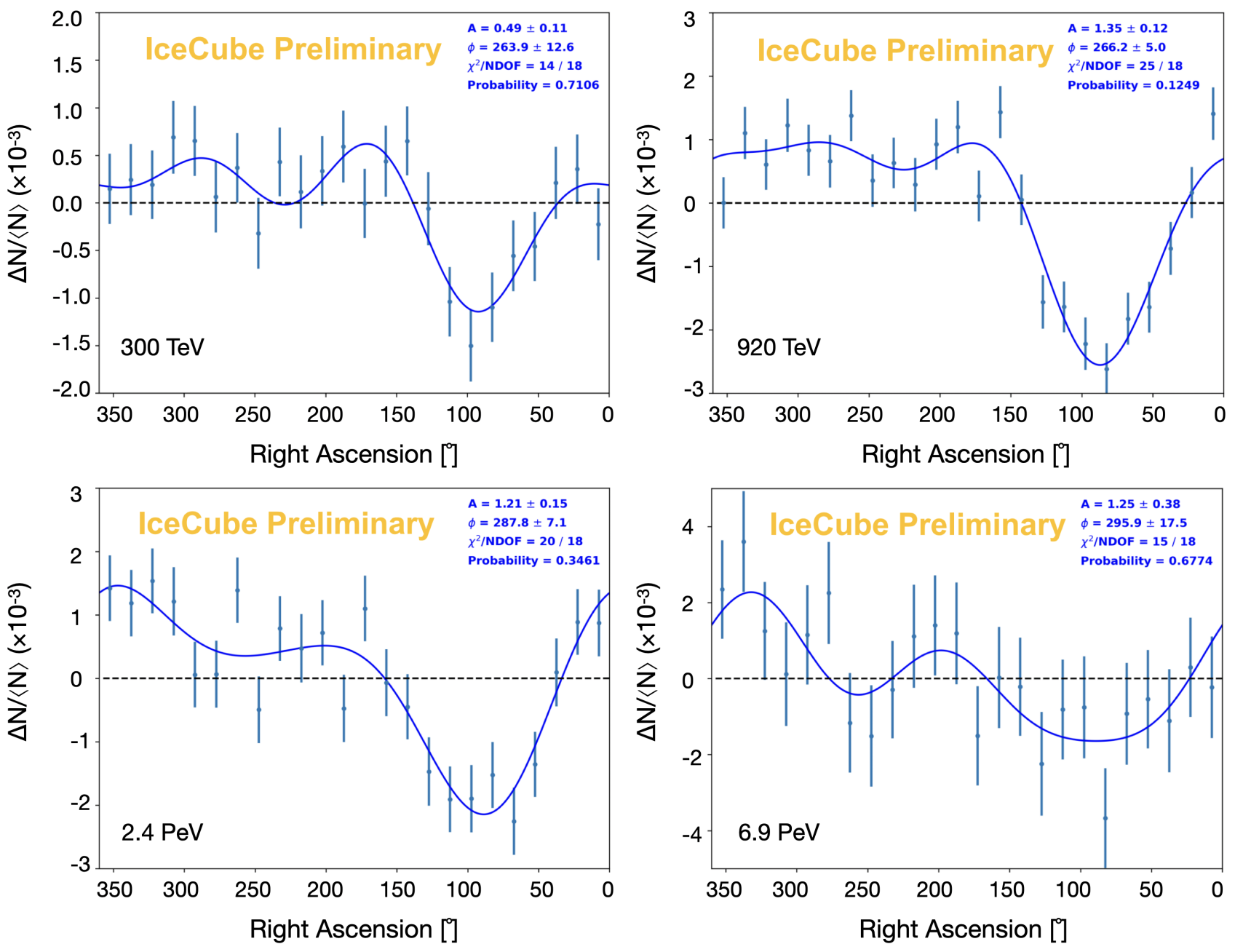

**Figure 4:** Relative intensity as a function of right ascension for four tiers with their corresponding median energies indicated.

## 4. Summary and Outlook

This work presents the analysis of eleven years of data collected by the IceTop air-shower array between May 2011 and May 2022, revealing an energy-dependent anisotropy in the arrival direction distribution of cosmic rays in the Southern Hemisphere. The amplitude of the anisotropy increases between 300 TeV and higher energies, rising from 0.49 at 300 TeV to 1.35 at 920 TeV, and remaining roughly constant at PeV energies (1.21 at 2.4 PeV and 1.25 at 6.9 PeV). Across all energy tiers, the large-scale relative intensity sky maps reveal a dominant deficit region in right ascension between approximately 60° and 120°, consistent with previous IceTop observations. To quantify the anisotropy, one-dimensional projections in right ascension were constructed, and a harmonic fit was applied, including up to the third-order component. The results show that a dipole alone is insufficient to describe the observed structure, and higher-order terms are necessary. While the location of the deficit remains relatively stable across energy bands, both the amplitude and width of the anisotropy increase with energy from 300 TeV to 6.9 PeV. These findings reaffirm the persistence and evolution of large-scale anisotropy features in the Southern Hemisphere cosmic-ray sky.

The extended dataset analyzed in this work enables a significant reduction in statistical uncertainties, allowing for more precise characterization of the anisotropy across a broad energy range. In particular, the inclusion of data from a stable, fully deployed detector—combined with improved analysis techniques and year-round coverage—enhances sensitivity to PeV-scale anisotropies and reduces systematic effects due to interference from Earth's orbital motion [30].

In future work, we plan to explore the solar dipole anisotropy and assess systematic uncertainties using the antisidereal projection method. The detection of cosmic-ray anisotropy with IceTop not only reinforces its role in studying arrival direction distributions, but also opens avenues for deeper investigations into the origin and propagation of cosmic rays. With its capability to measure both the energy spectrum and elemental composition of cosmic rays, IceTop is well-suited for future studies that link anisotropy patterns to changes in composition, energy-dependent diffusion processes, and potential nearby sources. These investigations will improve our understanding of the astrophysical mechanisms shaping the observed cosmic-ray sky.

## Full Author List: IceCube Collaboration

R. Abbasi[16], M. Ackermann[63], J. Adams[17], S. K. Agarwalla[39, a], G. Agrawal[16], J. A. Aguilar[10], M. Ahlers[21], J.M. Alameddine[22], S. Ali[35], N. M. Amin[43], K. Andeen[41], C. Argüelles[13], Y. Ashida[52], S. Athanasiadou[63], S. N. Axani[43], R. Babu[23], X. Bai[49], J. Baines-Holmes[39], A. Balagopal V.[39, 43], S. W. Barwick[29], S. Bash[26], E. Bastow[16], V. Basu[52], R. Bay[6], J. J. Beatty[19, 20], J. Becker Tjus[9, b], P. Behrens[1], J. Beise[61], C. Bellenghi[26], B. Benkel[63], S. BenZvi[51], D. Berley[18], E. Bernardini[47, c], D. Z. Besson[35], E. Blaufuss[18], L. Bloom[58], S. Blot[63], I. Bodo[39], F. Bontempo[30], J. Y. Book Motzkin[13], C. Boscolo Meneguolo[47, c], S. Böser[40], O. Botner[61], J. Böttcher[1], G. Bratrud[16], J. Braun[39], B. Brinson[4], Z. Brisson-Tsavoussis[32], R. T. Burley[2], D. Butterfield[39], M. A. Campana[48], K. Carloni[13], J. Carpio[33, 34], R. Chapagain[37], S. Chattopadhyay[39, a], N. Chau[10], Z. Chen[55], D. Chirkin[39], S. Choi[52], B. A. Clark[18], A. Coleman[61], P. Coleman[1], G. H. Collin[14], D. A. Coloma Borja[47], A. Connolly[19, 20], J. M. Conrad[14], R. Corley[52], D. F. Cowen[59, 60], C. De Clercq[11], J. J. DeLaunay[59], D. Delgado[13], T. Delmeulle[10], S. Deng[1], P. Desiati[39], K. D. de Vries[11], G. de Wasseige[36], T. DeYoung[23], J. C. Díaz-Vélez[39], S. DiKerby[23], M. Dittmer[42], A. Domi[25], L. Draper[52], L. Dueser[1], D. Durnford[24], K. Dutta[40], M. A. DuVernois[39], T. Ehrhardt[40], L. Eidenschink[26], A. Eimer[25], P. Eller[26], E. Ellinger[62], D. Elsässer[22], R. Engel[30, 31], H. Erpenbeck[39], W. Esmail[42], S. Eulig[13], J. Evans[18], P. A. Evenson[43], K. L. Fan[18], K. Fang[39], K. Farrag[15], A. R. Fazely[5], A. Fedynitch[57], N. Feigl[8], C. Finley[54], L. Fischer[63], D. Fox[59], A. Franckowiak[9], S. Fukami[63], P. Fürst[1], J. Gallagher[38], E. Ganster[1], A. Garcia[13], M. Garcia[43], G. Garg[39, a], E. Genton[13, 36], L. Gerhardt[7], A. Ghadimi[58], C. Glaser[61], T. Glüsenkamp[61], J. G. Gonzalez[43], S. Goswami[33, 34], A. Granados[23], D. Grant[12], S. J. Gray[18], S. Griffin[39], S. Griswold[51], K. M. Groth[21], D. Guevel[39], C. Günther[1], P. Gutjahr[22], C. Ha[53], C. Haack[25], A. Hallgren[61], L. Halve[1], F. Halzen[39], L. Hamacher[1], M. Ha Minh[26], M. Handt[1], K. Hanson[39], J. Hardin[14], A. A. Harnisch[23], P. Hatch[32], A. Haungs[30], J. Häußler[1], A. Hayes[37], K. Helbing[62], J. Hellrung[9], B. Henke[23], L. Hennig[25], F. Henningsen[12], L. Heuermann[1], R. Hewett[17], N. Heyer[61], S. Hickford[62], A. Hidvegi[54], C. Hill[15], G. C. Hill[2], R. Hmaid[15], K. D. Hoffman[18], D. Hooper[39], S. Hori[39], K. Hoshina[39, d], M. Hostert[13], W. Hou[30], T. Huber[30], K. Hultqvist[54], K. Hymon[22, 57], A. Ishihara[15], W. Iwakiri[15], M. Jacquart[21], S. Jain[39], O. Janik[25], M. Jansson[36], M. Jeong[52], M. Jin[13], N. Kamp[13], D. Kang[30], W. Kang[48], X. Kang[48], A. Kappes[42], L. Kardum[22], T. Karg[63], M. Karl[26], A. Karle[39], A. Katil[24], M. Kauer[39], J. L. Kelley[39], M. Khanal[52], A. Khatee Zathul[39], A. Kheirandish[33, 34], H. Kimku[53], J. Kiryluk[55], C. Klein[25], S. R. Klein[6, 7], Y. Kobayashi[15], A. Kochocki[23], R. Koirala[43], H. Kolanoski[8], T. Kontrimas[26], L. Köpke[40], C. Kopper[25], D. J. Koskinen[21], P. Koundal[43], M. Kowalski[8, 63], T. Kozynets[21], N. Krieger[9], J. Krishnamoorthi[39, a], T. Krishnan[13], K. Kruiswijk[36], E. Krupczak[23], A. Kumar[63], E. Kun[9], N. Kurahashi[48], N. Lad[63], C. Lagunas Gualda[26], L. Lallement Arnaud[10], M. Lamoureux[36], M. J. Larson[18], F. Lauber[62], J. P. Lazar[36], S. Lehrman[16], K. Leonard DeHolton[60], A. Leszczyńska[43], J. Liao[4], C. Lin[43], Y. T. Liu[60], M. Liubarska[24], C. Love[48], L. Lu[39], F. Lucarelli[27], W. Luszczak[19, 20], Y. Lyu[6, 7], J. Madsen[39], E. Magnus[11], K. B. M. Mahn[23], Y. Makino[39], E. Manao[26], S. Mancina[47, e], A. Mand[39], I. C. Mariş[10], S. Marka[45], Z. Marka[45], L. Marten[1], I. Martinez-Soler[13], R. Maruyama[44], J. Mauro[36], F. Mayhew[23], A. McClure[16], F. McNally[37], J. V. Mead[21], K. Meagher[39], S. Mechbal[63], A. Medina[20], M. Meier[15], Y. Merckx[11], L. Merten[9], J. Mitchell[5], L. Molchany[49], T. Montaruli[27], R. W. Moore[24], Y. Morii[15], A. Mosbrugger[25], M. Moulai[39], D. Mousadi[63], E. Moyaux[36], T. Mukherjee[30], R. Naab[63], M. Nakos[39], U. Naumann[62], J. Necker[63], L. Neste[54], M. Neumann[42], H. Niederhausen[23], M. U. Nisa[23], K. Noda[15], A. Noell[1], A. Novikov[43], A. Obertacke Pollmann[15], V. O'Dell[39], A. Olivas[18], R. Orsoe[26], J. Osborn[39], E. O'Sullivan[61], V. Palusova[40], H. Pandya[43], A. Parenti[10], N. Park[32], V. Parrish[23], E. N. Paudel[58], L. Paul[49], J. Paulson[16], C. Pérez de los Heros[61], T. Pernice[63], J. Peterson[39], M. Plum[49], A. Pontén[61], V. Poojyam[58], Y. Popovych[40], M. Prado Rodriguez[39], B. Pries[23], R. Procter-Murphy[18], G. T. Przybylski[7], L. Pyras[52], C. Raab[36], J. Rack-Helleis[40], N. Rad[63], M. Ravn[61], K. Rawlins[3], Z. Rechav[39], A. Rehman[43], I. Reistroffer[49], E. Resconi[26], S. Reusch[63], C. D. Rho[56], W. Rhode[22], L. Ricca[36], B. Riedel[39], A. Rifaie[62], E. J. Roberts[2], S. Robertson[6, 7], M. Rongen[25], A. Rosted[15], C. Rott[52], T. Ruhe[22], L. Ruohan[26], D. Ryckbosch[28], J. Saffer[31], D. Salazar-Gallegos[23], P. Sampathkumar[30], A. Sandrock[62], G. Sanger-Johnson[23], M. Santander[58], S. Sarkar[46], J. Savelberg[1], M. Scarnera[36], P. Schaile[26], M. Schaufel[1], H. Schieler[30], S. Schindler[25], L. Schlickmann[40], B. Schlüter[42], F. Schlüter[10], N. Schmeisser[62], T. Schmidt[18], F. G. Schröder[30, 43], L. Schumacher[25], S. Schwirn[1], S. Sclafani[18], D. Seckel[43], L. Seen[39], M. Seikh[35], S. Seunarine[50], P. A. Sevle Myhr[36], R. Shah[48], S. Shefali[31], N. Shimizu[15], B. Skrzypek[6], R. Snihur[39], J. Soedingrekso[22], A. Søgaard[21], D. Soldin[52], P. Soldin[1], G. Sommani[9], C. Spannfellner[26], G. M. Spiczak[50], C. Spiering[63], J. Stachurska[28], M. Stamatikos[20], T. Stanev[43], T. Stezelberger[7], T. Stürwald[62], T. Stuttard[21], G. W. Sullivan[18], J. Summers[16], I. Taboada[4], S. Ter-Antonyan[5], A. Terliuk[26], A. Thakuri[49], M. Thiesmeyer[39], W. G. Thompson[13], J. Thwaites[39], S. Tilav[43], K. Tollefson[23], S. Toscano[10], D. Tosi[39], A. Trettin[63], A. K. Upadhyay[39, a], K. Upshaw[5], A. Vaidyanathan[41], N. Valtonen-Mattila[9, 61], J. Valverde[41], J. Vandenbroucke[39], T. van Eeden[63], N. van Eijndhoven[11], L. van Rootselaar[22], J. van Santen[63], F. J. Vara Carbonell[42], F. Varsi[31], M. Venugopal[30], M. Vereecken[36], S. Vergara Carrasco[17], S. Verpoest[43], D. Veske[45], A. Vijai[18], J. Villarreal[14], C. Walck[54], A. Wang[4], E. Warrick[58], C. Weaver[23], P. Weigel[14], A. Weindl[30], J. Weldert[40], A. Y. Wen[13], C. Wendt[39], J. Werthebach[22], M. Weyrauch[30], N. Whitehorn[23], C. H. Wiebusch[1], D. R. Williams[58], L. Witthaus[22], M. Wolf[26], G. Wrede[25], X. W. Xu[5], J. P. Yañez[24], Y. Yao[39], E. Yildizci[39], S. Yoshida[15], R. Young[35], F. Yu[13], S. Yu[52], T. Yuan[39], A. Zegarelli[9], S. Zhang[23], Z. Zhang[55], P. Zhelnin[13], P. Zilberman[39]

[1] III. Physikalisches Institut, RWTH Aachen University, D-52056 Aachen, Germany
[2] Department of Physics, University of Adelaide, Adelaide, 5005, Australia
[3] Dept. of Physics and Astronomy, University of Alaska Anchorage, 3211 Providence Dr., Anchorage, AK 99508, USA
[4] School of Physics and Center for Relativistic Astrophysics, Georgia Institute of Technology, Atlanta, GA 30332, USA
[5] Dept. of Physics, Southern University, Baton Rouge, LA 70813, USA
[6] Dept. of Physics, University of California, Berkeley, CA 94720, USA
[7] Lawrence Berkeley National Laboratory, Berkeley, CA 94720, USA
[8] Institut für Physik, Humboldt-Universität zu Berlin, D-12489 Berlin, Germany
[9] Fakultät für Physik & Astronomie, Ruhr-Universität Bochum, D-44780 Bochum, Germany

[10] Université Libre de Bruxelles, Science Faculty CP230, B-1050 Brussels, Belgium
[11] Vrije Universiteit Brussel (VUB), Dienst ELEM, B-1050 Brussels, Belgium
[12] Dept. of Physics, Simon Fraser University, Burnaby, BC V5A 1S6, Canada
[13] Department of Physics and Laboratory for Particle Physics and Cosmology, Harvard University, Cambridge, MA 02138, USA
[14] Dept. of Physics, Massachusetts Institute of Technology, Cambridge, MA 02139, USA
[15] Dept. of Physics and The International Center for Hadron Astrophysics, Chiba University, Chiba 263-8522, Japan
[16] Department of Physics, Loyola University Chicago, Chicago, IL 60660, USA
[17] Dept. of Physics and Astronomy, University of Canterbury, Private Bag 4800, Christchurch, New Zealand
[18] Dept. of Physics, University of Maryland, College Park, MD 20742, USA
[19] Dept. of Astronomy, Ohio State University, Columbus, OH 43210, USA
[20] Dept. of Physics and Center for Cosmology and Astro-Particle Physics, Ohio State University, Columbus, OH 43210, USA
[21] Niels Bohr Institute, University of Copenhagen, DK-2100 Copenhagen, Denmark
[22] Dept. of Physics, TU Dortmund University, D-44221 Dortmund, Germany
[23] Dept. of Physics and Astronomy, Michigan State University, East Lansing, MI 48824, USA
[24] Dept. of Physics, University of Alberta, Edmonton, Alberta, T6G 2E1, Canada
[25] Erlangen Centre for Astroparticle Physics, Friedrich-Alexander-Universität Erlangen-Nürnberg, D-91058 Erlangen, Germany
[26] Physik-department, Technische Universität München, D-85748 Garching, Germany
[27] Département de physique nucléaire et corpusculaire, Université de Genève, CH-1211 Genève, Switzerland
[28] Dept. of Physics and Astronomy, University of Gent, B-9000 Gent, Belgium
[29] Dept. of Physics and Astronomy, University of California, Irvine, CA 92697, USA
[30] Karlsruhe Institute of Technology, Institute for Astroparticle Physics, D-76021 Karlsruhe, Germany
[31] Karlsruhe Institute of Technology, Institute of Experimental Particle Physics, D-76021 Karlsruhe, Germany
[32] Dept. of Physics, Engineering Physics, and Astronomy, Queen's University, Kingston, ON K7L 3N6, Canada
[33] Department of Physics & Astronomy, University of Nevada, Las Vegas, NV 89154, USA
[34] Nevada Center for Astrophysics, University of Nevada, Las Vegas, NV 89154, USA
[35] Dept. of Physics and Astronomy, University of Kansas, Lawrence, KS 66045, USA
[36] Centre for Cosmology, Particle Physics and Phenomenology - CP3, Université catholique de Louvain, Louvain-la-Neuve, Belgium
[37] Department of Physics, Mercer University, Macon, GA 31207-0001, USA
[38] Dept. of Astronomy, University of Wisconsin—Madison, Madison, WI 53706, USA
[39] Dept. of Physics and Wisconsin IceCube Particle Astrophysics Center, University of Wisconsin—Madison, Madison, WI 53706, USA
[40] Institute of Physics, University of Mainz, Staudinger Weg 7, D-55099 Mainz, Germany
[41] Department of Physics, Marquette University, Milwaukee, WI 53201, USA
[42] Institut für Kernphysik, Universität Münster, D-48149 Münster, Germany
[43] Bartol Research Institute and Dept. of Physics and Astronomy, University of Delaware, Newark, DE 19716, USA
[44] Dept. of Physics, Yale University, New Haven, CT 06520, USA
[45] Columbia Astrophysics and Nevis Laboratories, Columbia University, New York, NY 10027, USA
[46] Dept. of Physics, University of Oxford, Parks Road, Oxford OX1 3PU, United Kingdom
[47] Dipartimento di Fisica e Astronomia Galileo Galilei, Università Degli Studi di Padova, I-35122 Padova PD, Italy
[48] Dept. of Physics, Drexel University, 3141 Chestnut Street, Philadelphia, PA 19104, USA
[49] Physics Department, South Dakota School of Mines and Technology, Rapid City, SD 57701, USA
[50] Dept. of Physics, University of Wisconsin, River Falls, WI 54022, USA
[51] Dept. of Physics and Astronomy, University of Rochester, Rochester, NY 14627, USA
[52] Department of Physics and Astronomy, University of Utah, Salt Lake City, UT 84112, USA
[53] Dept. of Physics, Chung-Ang University, Seoul 06974, Republic of Korea
[54] Oskar Klein Centre and Dept. of Physics, Stockholm University, SE-10691 Stockholm, Sweden
[55] Dept. of Physics and Astronomy, Stony Brook University, Stony Brook, NY 11794-3800, USA
[56] Dept. of Physics, Sungkyunkwan University, Suwon 16419, Republic of Korea
[57] Institute of Physics, Academia Sinica, Taipei, 11529, Taiwan
[58] Dept. of Physics and Astronomy, University of Alabama, Tuscaloosa, AL 35487, USA
[59] Dept. of Astronomy and Astrophysics, Pennsylvania State University, University Park, PA 16802, USA
[60] Dept. of Physics, Pennsylvania State University, University Park, PA 16802, USA
[61] Dept. of Physics and Astronomy, Uppsala University, Box 516, SE-75120 Uppsala, Sweden
[62] Dept. of Physics, University of Wuppertal, D-42119 Wuppertal, Germany
[63] Deutsches Elektronen-Synchrotron DESY, Platanenallee 6, D-15738 Zeuthen, Germany
[a] also at Institute of Physics, Sachivalaya Marg, Sainik School Post, Bhubaneswar 751005, India
[b] also at Department of Space, Earth and Environment, Chalmers University of Technology, 412 96 Gothenburg, Sweden
[c] also at INFN Padova, I-35131 Padova, Italy
[d] also at Earthquake Research Institute, University of Tokyo, Bunkyo, Tokyo 113-0032, Japan
[e] now at INFN Padova, I-35131 Padova, Italy

## Acknowledgments

The authors gratefully acknowledge the support from the following agencies and institutions: USA – U.S. National Science Foundation-Office of Polar Programs, U.S. National Science Foundation-Physics Division, U.S. National Science Foundation-EPSCoR, U.S. National Science Foundation-Office of Advanced Cyberinfrastructure, Wisconsin Alumni Research Foundation, Center for High Throughput Computing (CHTC) at the University of Wisconsin–Madison, Open Science Grid (OSG), Partnership to Advance Throughput Computing (PATh), Advanced Cyberinfrastructure Coordination Ecosystem: Services & Support (ACCESS), Frontera and Ranch computing project at the Texas Advanced Computing Center, U.S. Department of Energy-National Energy Research Scientific Computing Center, Particle astrophysics research computing center at the University of Maryland, Institute for Cyber-Enabled Research at Michigan State University, Astroparticle physics computational facility at Marquette University, NVIDIA Corporation, and Google Cloud Platform; Belgium – Funds for Scientific Research (FRS-FNRS and FWO), FWO Odysseus and Big Science programmes, and Belgian Federal Science Policy Office (Belspo); Germany – Bundesministerium für Forschung, Technologie und Raumfahrt (BMFTR), Deutsche Forschungsgemeinschaft (DFG), Helmholtz Alliance for Astroparticle Physics (HAP), Initiative and Networking Fund of the Helmholtz Association, Deutsches Elektronen Synchrotron (DESY), and High Performance Computing cluster of the RWTH Aachen; Sweden – Swedish Research Council, Swedish Polar Research Secretariat, Swedish National Infrastructure for Computing (SNIC), and Knut and Alice Wallenberg Foundation; European Union – EGI Advanced Computing for research; Australia – Australian Research Council; Canada – Natural Sciences and Engineering Research Council of Canada, Calcul Québec, Compute Ontario, Canada Foundation for Innovation, WestGrid, and Digital Research Alliance of Canada; Denmark – Villum Fonden, Carlsberg Foundation, and European Commission; New Zealand – Marsden Fund; Japan – Japan Society for Promotion of Science (JSPS) and Institute for Global Prominent Research (IGPR) of Chiba University; Korea – National Research Foundation of Korea (NRF); Switzerland – Swiss National Science Foundation (SNSF).